\documentclass{article}

\usepackage{amsmath}
\usepackage{amssymb}
\usepackage{authblk}
\usepackage{physics}
\usepackage[numbers, sort]{natbib}
\usepackage{hyperref}
\usepackage{comment}
\usepackage{color}
\usepackage{graphicx}
\usepackage{fancyhdr}
\usepackage{geometry}
\usepackage{caption}
\usepackage{subcaption}
\usepackage{float}
\newcommand*{\sectioncolor}{black}
\newcommand*{\sectionformat}{\centering\color{\sectioncolor}}

\author[1, 2]{\small Calum Maitland*}
\author[2]{\small Daniele Faccio}
\author[1]{\small Fabio Biancalana}

\affil[1]{\footnotesize Institute of Photonics and Quantum Sciences, Heriot-Watt University, Edinburgh EH14 4AS, UK}
\affil[2]{\footnotesize School of Physics and Astronomy, Kelvin Building, University of Glasgow, Glasgow G12 8QQ, UK}
\affil[ ]{\textit{*cm350@hw.ac.uk}}

\date{}

\begin{document}

\title{Modulation Instability of Discrete Angular Momentum in Coupled Fiber Rings}

\maketitle

\begin{abstract}
We present an analysis of temporal modulation instability in a ring array of coupled optical fibers. Continuous-wave signals are shown to be unstable to perturbations carrying discrete angular momenta, both for normal and anomalous group velocity dispersion. We find the frequency spectrum of modulation instability is different for each perturbation angular momentum and depends strongly on the coupling strength between fibers in the ring. Twisting the ring array also allows the  frequency spectra to be tuned through the induced tunnelling Peierls phase.
\end{abstract}

\bigskip
\section*{\sectionformat I. Introduction}

\begin{figure}[htbp] 
  \centering
  \includegraphics[width=8cm]{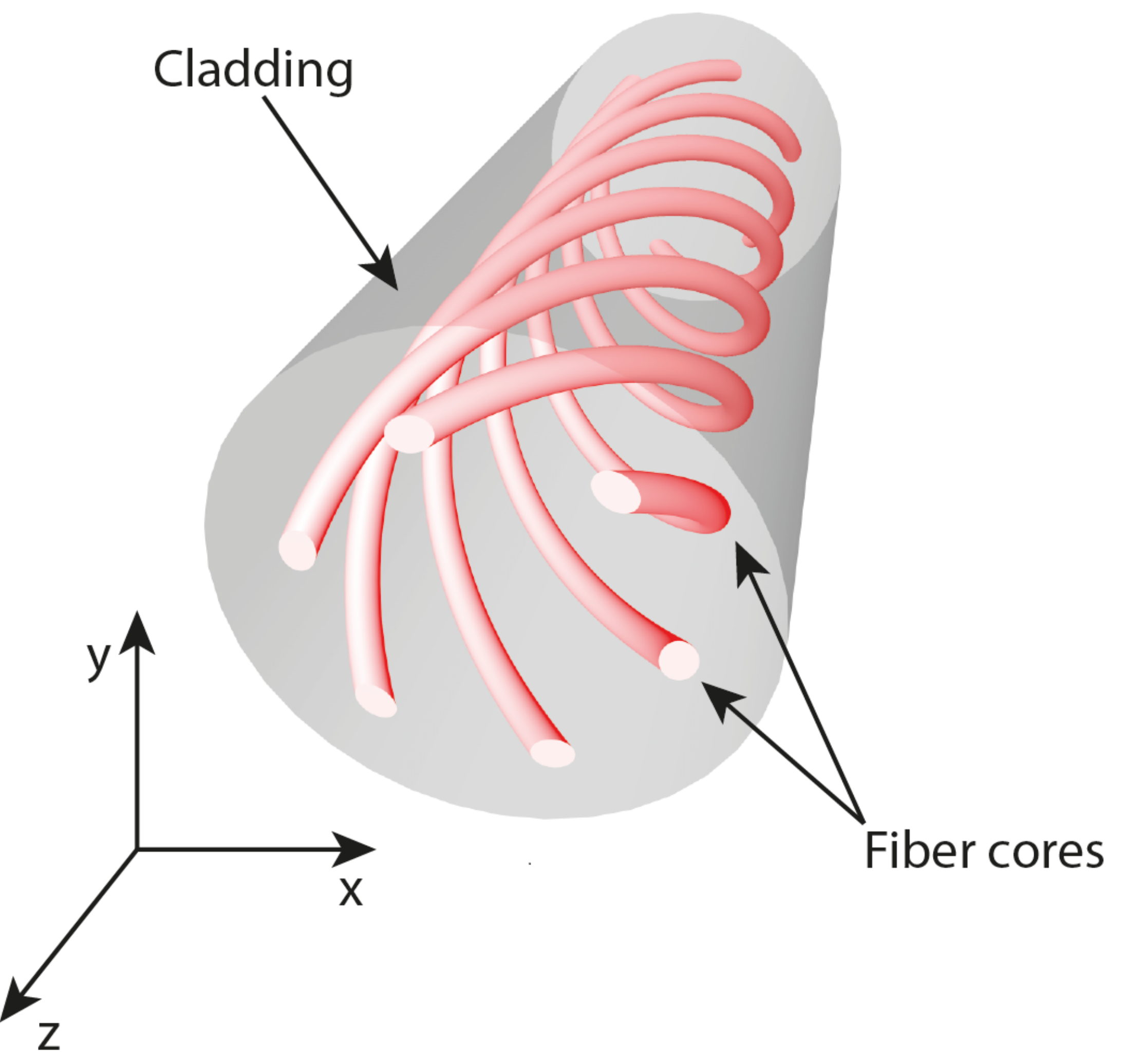}
\caption{\small{Six core fiber ring over a single twist period $\Lambda$. Light couples between cores through the overlap of the evanescent fields of their fundamental guided modes (not shown).}} \label{fig: fiberring}
\end{figure}

Modulation instability (MI) of plane waves in anomalously-dispersive optical fibers with a self-focusing Kerr nonlinearity is one of the most well-known phenomena in nonlinear optics \cite{Agrawal2001}. Continuous-wave signals subject to MI break up into a train of pulses as fluctuations grow through the nonlinearity, resulting in a characteristic spectrum of a pair of symmetric sidebands about the signal's original frequency. It has been observed in higher dimensional systems, in particular as an azimuthal instability of optical vortices in continuous self-focusing quadratic \cite{Petrov1998}, Kerr \cite{Kruglov1992, Vuong2006}, saturable \cite{Silahli2015, Walasik2017} and defocusing Kerr \cite{Law1994} nonlinear media. An understanding of MI in a system can be seen as the basis for exploring supercontinuum generation, soliton dynamics and many other nonlinear phenomena therein \cite{Saleh2012, Saleh2016}. Circular arrays of coupled optical fibers have been shown to support supermodes carrying angular momentum as discrete optical vortices \cite{Ferrando2005, Alexeyev2009}. By twisting these arrays along their propagation axis, discrete diffraction cancellation and other ways to manipulate optical tunnelling have been discovered \cite{Ornigotti2007, Longhi2007, Longhi2007b}. Some nonlinear optical properties of fiber rings have been examined previously, particularly in the context of $\mathcal{PT}$-symmetry breaking \cite{Castro2016}, optical switching \cite{Tofighi2013} and  the stability of modes in fibers with only a few cores \cite{Li2016}. Here we show that plane wave supermodes of fiber arrays as shown in figure \ref{fig: fiberring} can be unstable in the presence of perturbations and that the gain spectra of these perturbations depends on their angular momenta. Until now, this temporal MI of discrete angular momentum signals has been little explored compared to the well-known azimuthal instability pointed out above. The generation of angular momentum modes over a wide range of frequencies is of practical interest for multiplexing in optical communications \cite{Bozinovic2013, Willner2015, Bruning2016}. We find a previously unknown, rich spectral structure in their instability gain spectra. A Peierls phase introduced by coiling the ring around its propagation axis adds a further degree of depth and potential control to these spectra, but it is by no means necessary to observe this novel MI.

We first outline our derivation of MI in a general fiber ring array with $N$ cores. Then we exhibit some calculated MI gain spectra for perturbations carrying angular momentum in straight six-core rings. Lastly we examine how twisting the array can introduce (or suppress) unstable modes, depending on the coupling between neighbouring cores in the ring.

\bigskip
\section*{\sectionformat II. Instability Model}

We model light propagation along the spatial axis $z$ of the $N$-core fiber array over time $t$ with $N$ coupled (1+1) dimensional nonlinear Schr\"{o}dinger equations; the equation for the electric field $E_n$ in the $n^{th}$ fiber is

\begin{equation} \label{eq: coupledNLSE}
i \partial_z E_n = \frac{\beta_2}{2} {\partial_t}^2 E_n - \gamma {|E_n|}^2 E_n  - \Delta \left(\exp(-i \phi) E_{n+1} + \exp(i \phi) E_{n-1} \right)
\end{equation}

in which $\beta_2$ is the group velocity dispersion, $\gamma$ is the nonlinear coefficient, $\Delta$ the coupling strength between nearest-neighbour fiber cores and $+(-)\phi$ the Peierl's phase acquired by photons tunnelling between cores in the direction of (against) the array twist. This Peierls phase is analogous to the phase aquired by charged particles travelling along a magnetic vector potential through the Aharanov-Bohm effect \cite{Longhi2014}. $\Delta$ is proportional to the overlap between guided modes in adjacent cores \cite{Longhi2007, Longhi2007b}; as such it is extremely sensitive to the distance separating cores and the extent of their fundamental modes' evanescent fields. Careful engineering of these degrees of freedom should allow $\Delta$ to be tuned over several orders of magnitude. Here we have negelected fiber losses for simplicity's sake, but their inclusion in this model would be straightforward. Following the standard MI derivation in Agrawal \cite{Agrawal2001}, we consider a plane-wave solution of \eqref{eq: coupledNLSE} with optical power $P_0$ incident in each core (indexed by the integer $n$ running from $1$ to $N$) which carries angular momentum with winding number $m$

\begin{equation} \label{eq: planewavesols}
\overline{E}_n = \sqrt{P_0} \exp \left(i 2 \pi m n/N +  i k_0 z \right)
\end{equation}

as a pump with wavenumber $k_0 =  \gamma P_0 + 2 \Delta \cos\left( 2 \pi m /N - \phi \right)$. This is perturbed by a pair of weak excitations, with (possibly complex) wavenumber $K$ and frequency detuning $\Omega$ from the pump, coupled through the nonlinearity

\begin{equation} \label{eq: Agrsigidlansatz}
a_n(z,t) = a_S \exp \left(i \left( K z + 2 \pi l n/N - \Omega t \right) \right) + a_I \exp \left(-i \left( K z + 2 \pi (l-2m) n/N - \Omega t \right) \right)
\end{equation}

which are denoted as signal ($\propto a_S$) and idler ($\propto a_I$). The finite number of cores restricts the signal winding number $l$ to a limited set of integers, $l \in [-N/2+1, N/2]$ for $N$ even or $l \in [-(N-1)/2, (N-1)/2]$ for $N$ odd. These perturbations will grow exponentially if $K$ has an imaginary component, which requires $\Omega$ to lie between two critical frequencies, $\Omega_{C1}$ and $\Omega_{C2}$:

\begin{equation} \label{eq: OmegaC}
\begin{split}
\Omega_{C1} &= \pm \Re\left( \sqrt{-\frac{2}{\beta_2}  \left( 2 \gamma P_0 + \Delta_{l, m} + \Delta_{2m-l, m} \right) } \right)\\
\Omega_{C2} &= \pm \Re \left( \sqrt{-\frac{2}{\beta_2} \left( \Delta_{l, m} + \Delta_{2m-l, m} \right)} \right)
\end{split}
\end{equation}

If this holds, then the perturbations grow exponentially in power at a rate

\begin{equation} \label{eq: gain}
G(\Omega, \Delta) =  \Re\left( \sqrt{ 4 \gamma^2 {P_0}^2 - {\left( \beta_2 \Omega^2 + 2 \gamma P_0 + 2\Delta_{l, m} + 2\Delta_{2m-l, m} \right) }^2 } \right)
\end{equation}

where we have defined $\Delta_{x, m} \equiv \Delta \left( \cos \left( 2 \pi x /N -\phi\right) - \cos \left( 2 \pi m /N -\phi\right) \right)$. To clarify, the gain rate $G(\Omega, \Delta) \equiv 2 \Im \left(K\right)$, meaning the perturbation intensity increases with propagation distance $z$ as ${|a_n|}^2 \propto \exp \left( G z \right)$ \cite{Agrawal2001}.  The frequency where this gain is maximised is given by 

\begin{equation} \label{eq: Omegamax}
\Omega_{max} =  \pm\Re\left( \sqrt{ -\frac{2}{\beta_2}\left(\gamma P_0 + \Delta_{l, m} + \Delta_{2m-l, m}\right) } \right)
\end{equation}

\bigskip
\section*{\sectionformat III. Straight Array Gain Spectra}

\begin{figure}[htbp] 
  \centering
  \includegraphics[width=12cm]{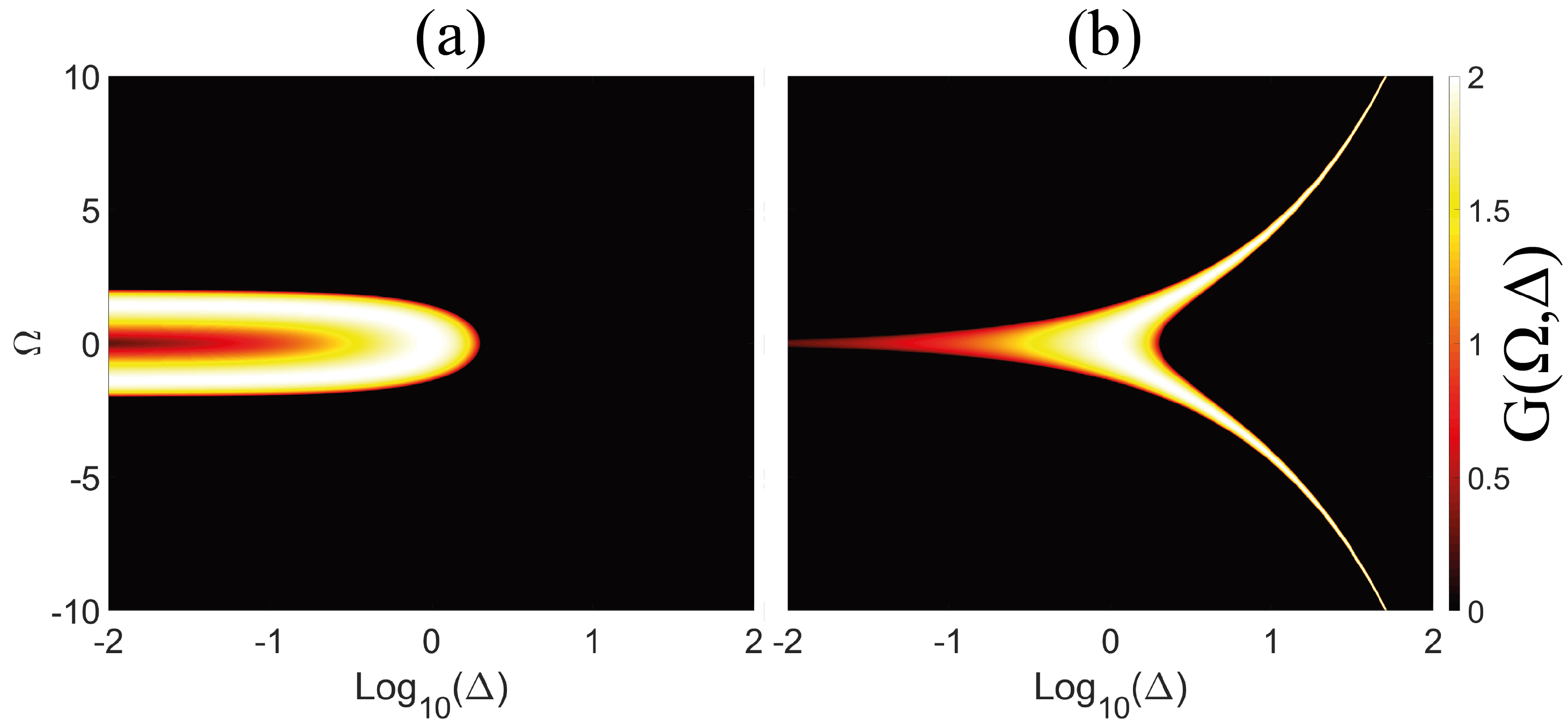}
\caption{\small{Gain spectra for perturbations carrying angular momentum $|l|=1$ as a function of the detuning $\Omega$ and the fiber coupling strength $\Delta$, in an untwisted array ($\phi=0$). \textbf{(a)} Anomalous GVD $\beta_2=-1$; \textbf{(b)} Normal GVD $\beta_2=1$. Here $\gamma P_0 = 1$.}} \label{spectra}
\end{figure}

Example plots of gain spectra in an untwisted ($\phi=0$) six-core fiber ring are shown in figure \ref{spectra}, given pump and signal angular momentum $m=0$ and $|l|=1$ respectively.

In contrast to standard MI, here we see MI can occur for both normal and anomalous GVD. If $\beta_2 < 0$, pertubations with non-zero $l$ see suppressed instability above a threshold coupling strength. With $\beta_2 > 0$, gain initially grows stronger with increasing coupling, until it forks into two sidebands. Higher order angular momenta experience similar instabilities, with their sidebands narrowed and shifted to higher $|\Omega|$. Whenever $l=m$, $\Delta_{l, m}=\Delta_{2m-l, m}=0$ meaning that the gain is insensitive to the coupling parameters and the standard results for MI in a single fiber apply. In general, the peak gain is capped at $G_{\max} = 2\gamma P_0$. 

To verify these analytical results, we numerically simulate propagation of a plane wave of power $P_0$ and no angular momentum ($m=0$) in a strongly-coupled straight six-fiber array with the full coupled NLSE \eqref{eq: coupledNLSE}. At the start of propagation $z=0$ we add complex noise $f_n(t)$ to each fiber with average power $\approx 10^{-8} P_0$ to mimic fluctuations from which spontaneous MI may emerge. We decompose the resulting field into its angular momentum $l$ and frequency $\Omega$ spectrum via the projection

\begin{equation} \label{eq: spectrumdecomp}
\tilde{E}_l(\Omega, z) = \sum_{n=1}^{N} \exp\left(i \frac{2 \pi l n}{N}\right) \int_{-\infty}^{\infty} dt \exp\left(- i \Omega t\right) E_n(t, z)
\end{equation} 

\begin{figure}[htbp] 
  \centering
  \includegraphics[width=12cm]{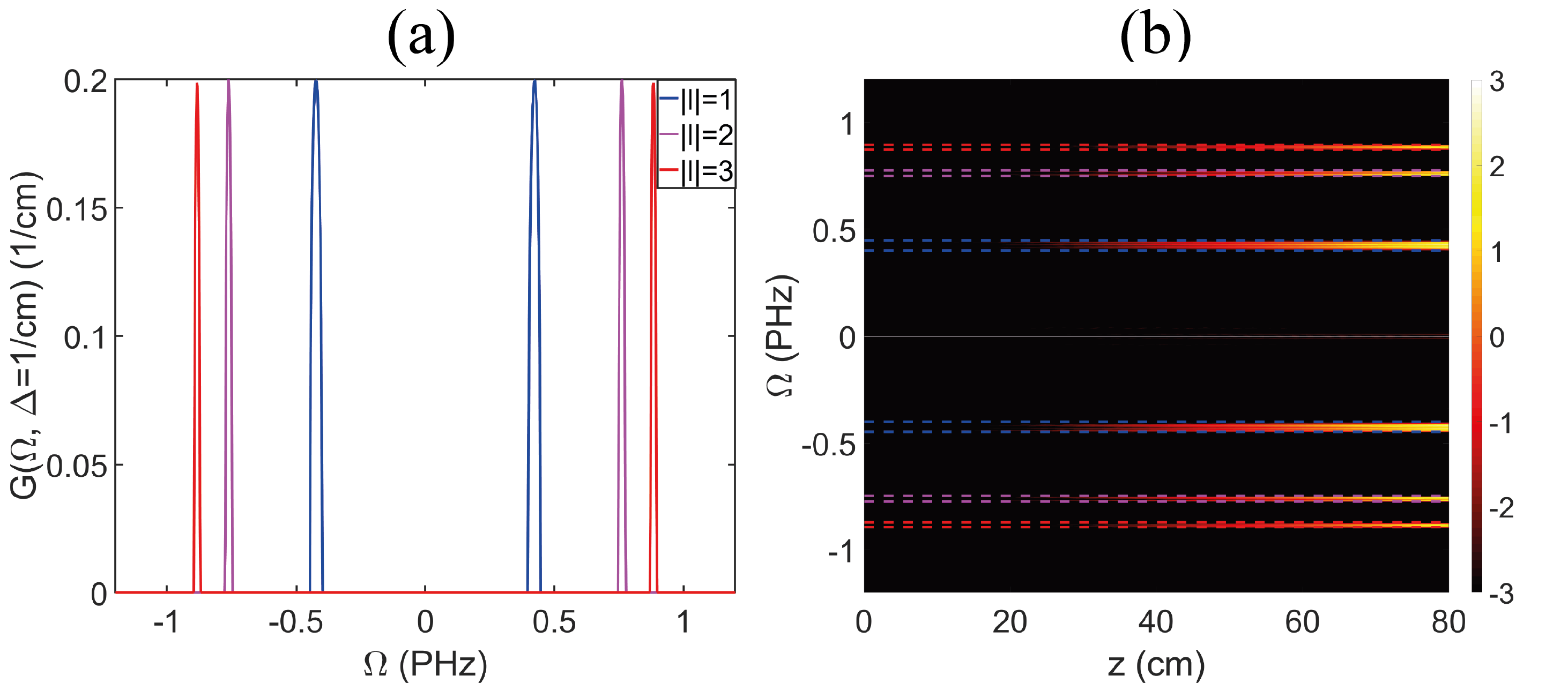}
\caption{\small{\textbf{(a)} Predicted MI gain for perturbations carrying different angular momenta $|l|=1,2,3$ in an homogeneously pumped six-core fiber ring with $\beta_2=1$ps$^2/$km, $\Delta=1/$cm, $\gamma P_0=0.1/$cm \textbf{(b)} Spectrum $|E_1(\Omega, z)|^2$ in a single core of the ring from a numerical simulation of eq. \eqref{eq: coupledNLSE}. Blue, magenta and red dashed lines indicate the MI critical frequencies for $|l|=1,2,3$, verifying the accuracy of eq. \eqref{eq: gain}. }} \label{simulation}
\end{figure}

Results are shown in figure \ref{simulation}, which compares the gain sidebands for $|l|=1,2,3$  \textbf{(a)} predicted by eq. \eqref{eq: gain} for a normally-dispersive, untwisted six-core fiber ring with the spectra in a single core \textbf{(b)} obtained by numerically integrating eq. \eqref{eq: coupledNLSE}. 

We can make some general observations in the case of normal dispersion; in the strong coupling limit $\Delta > \gamma P_0$, increasing $\Delta$  simultaneously shifts $\Omega_{max}$ and shrinks the bandwidth $|\Omega_{C1}-\Omega_{C2}|$, 
while in the weak coupling regime $\Delta < \gamma P_0$, $\Omega_{max}=0$ and decreasing $\Delta$ reduces both the bandwidth and magnitude of the gain. 

\section*{\sectionformat IV. Twisted Array Gain Spectra}

Twisting the fiber ring adds an additional degree of freedom to the MI gain spectrum by way of the Peierls phase $\phi$, which is related to the fiber twist period $\Lambda$ as \cite{Longhi2016}

\begin{equation} \label{eq: twistphase}
\phi =  \frac{8 \pi^3 n_s {r_0}^2}{N \lambda \Lambda}
\end{equation}

where $n_s$ is the substrate refractive index, $r_0$ approximately the ring radius,  $N$ the number of cores and $\lambda$ the central pump wavelength. When the twisted array is pumped with $m=0$ light, the gain spectrum is simple to characterise as only the real part $\Delta \cos(\phi)$ of the complex tunnelling coefficient $\Delta \exp\left(i \phi \right)$ is relevant. This is not the case when $m \neq 0$, however the same spectral structure emerges, centred around a different $\phi \neq 0$. 
We can calculate the twist phase about which the gain spectra are centred for  particular $m$ by finding $\phi_0$ which gives the greatest $\Omega_{max}$:

\begin{equation} \label{eq: phi0}
\phi_0 = 
\begin{cases}
\frac{2 \pi m}{N}, & \beta_2 > 0\\
\frac{2 \pi m}{N} + \pi, & \beta_2 < 0\\
\end{cases}
\end{equation}

\begin{figure} 
  \centering
  \includegraphics[width=12cm]{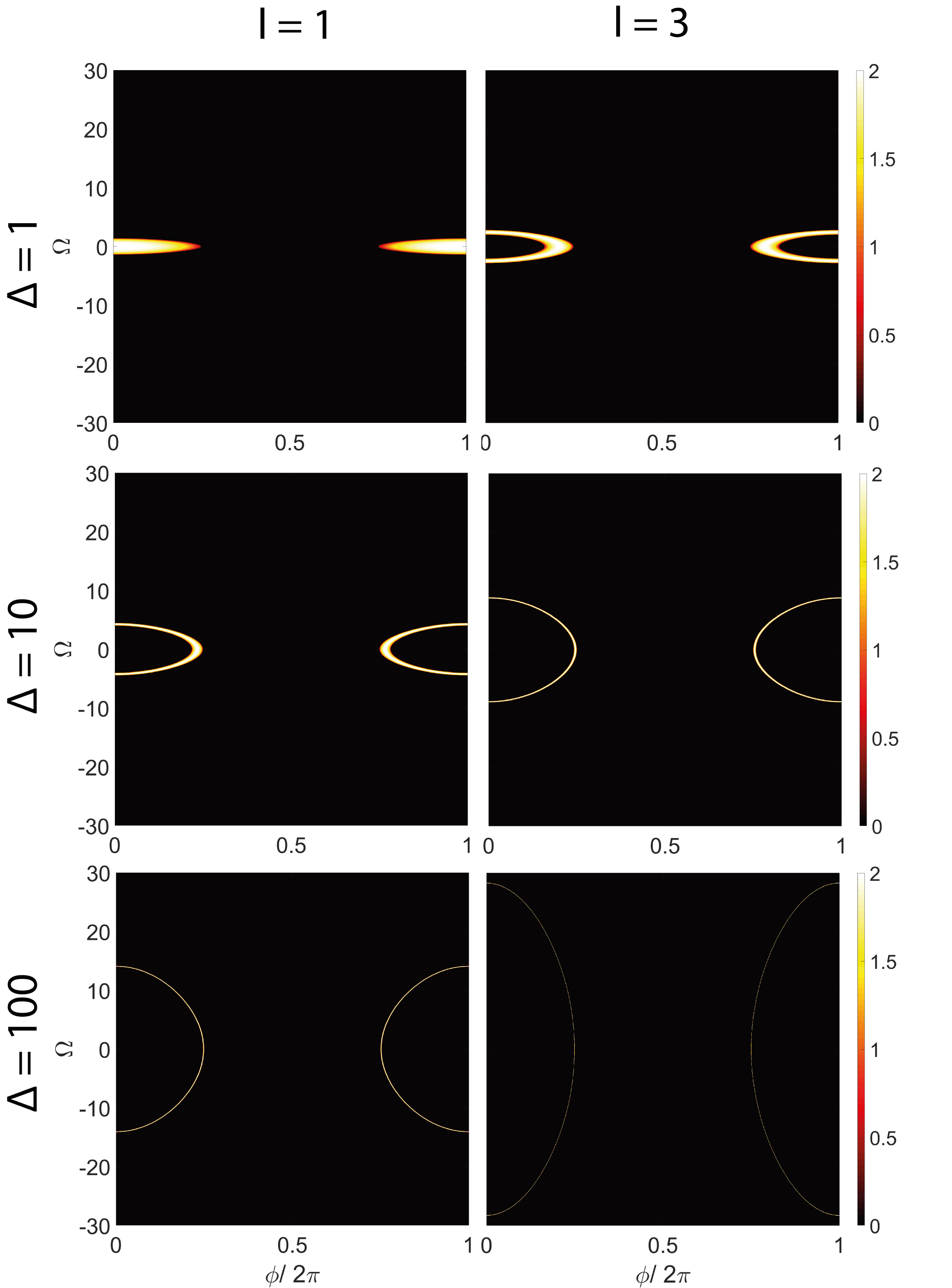}
 
\caption{\small{Gain spectra $G(\Omega, \phi)$ at various coupling strengths $\Delta$ for perturbations carrying angular momentum $l=\pm1$ (left column) and $l=\pm3$ (right column), given normal dispersion $\beta_2=1$ and a $m=0$ pump.}} \label{GainSpec_NormM=0}
\end{figure}
\begin{figure}[t]
  \centering
  \includegraphics[width=12cm]{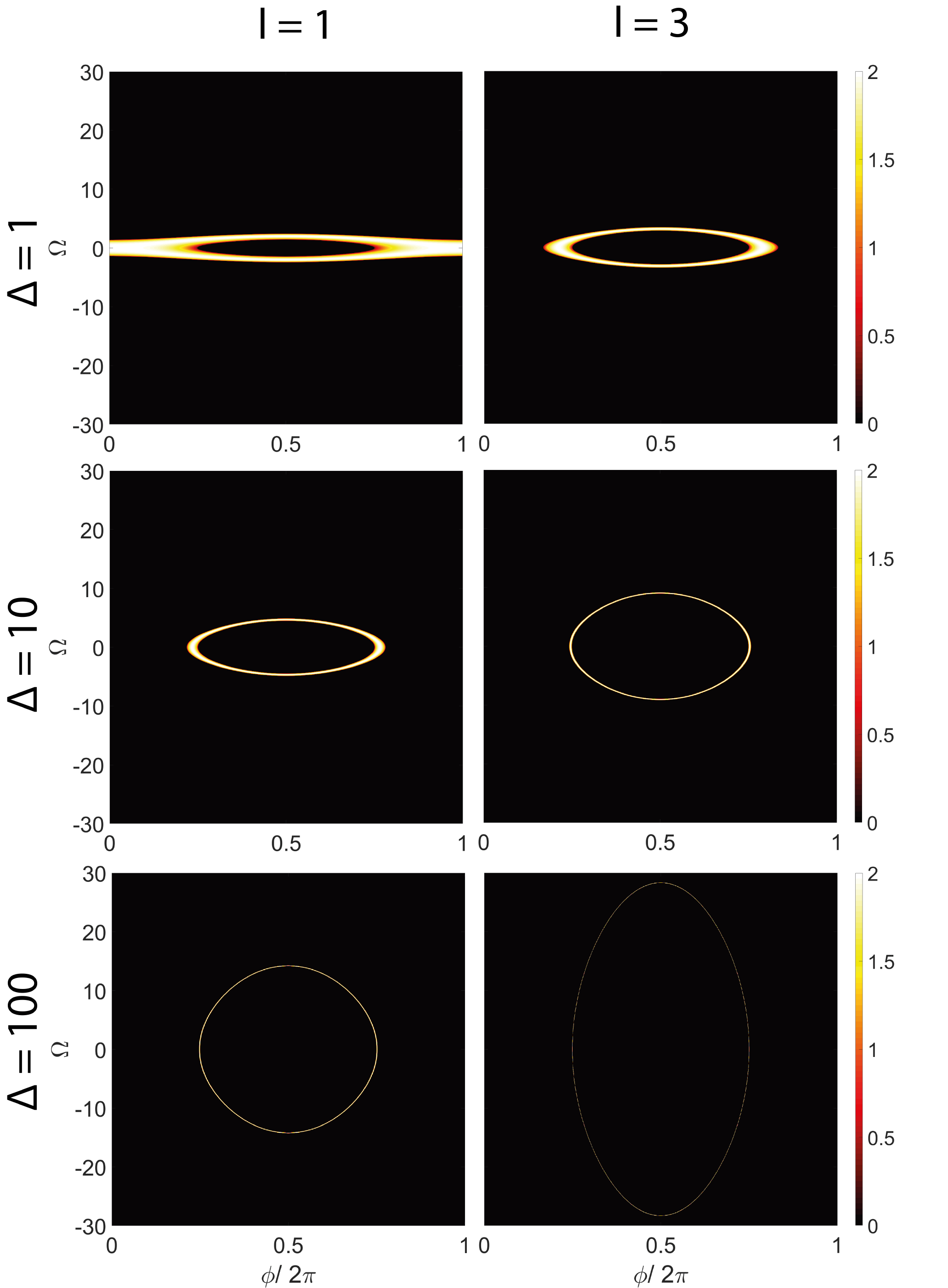}
\caption{\small{Gain spectra $G(\Omega, \phi)$ at various coupling strengths $\Delta$ for perturbations carrying angular momentum $l=\pm1$ (left column) and $l=\pm3$ (right column), given normal dispersion $\beta_2=1$ and a $m=0$ pump. }} \label{GainSpec_AnomM=0}
\end{figure}

Otherwise the spectra with common winding number differences $|l-m|$ are identical for varying $m$. Figures \ref{GainSpec_NormM=0} and \ref{GainSpec_AnomM=0} show the gain spectrum's dependence on $\phi$ for $|l-m|=1, 3$ in a six-core array pumped with light carrying no angular momentum $m=0$, for normal and anomalous dispersion respectively. As might be expected, the effect of varying $\phi$ is more noticeable at higher coupling strengths. Comparing normal and anomalous dispersion results in the strong coupling limit $\Delta=100$, we see very similar spectral structures centred around different $\phi_0$ as described by \eqref{eq: phi0}. As seen in the untwisted ring, a large coupling also leads to higher peak frequencies $\Omega_{max}$ and a reduced bandwidth. It is apparent from the results that MI can be suppressed by choosing $\phi = \phi_0 \pm \pi$, given any $\Delta$ for $\beta_2 > 0$ or a sufficiently big $\Delta \geq \gamma P_0 /2$ for $\beta_2 < 0$. 
In practice $\phi$ will be limited to small values, as strong twisting will also reduce $\Delta$ since it will push light within cores to their outer edges through an effective `centrifugal' force \cite{Longhi2007b}. However, it is not impossible that these detrimental effects may be countered by fabrication innovations, allowing for larger $\phi$ to be achieved through twisting. Alternative schemes for realising the Peierls phase may also be possible. As noted earlier, it can be equivalently described due to a synthetic magnetic field for photons, oriented along the $z$ axis of the fiber ring. Artificial gauge fields for light is a research area making steady progress at present \cite{Fang2012, Longhi2014, Westerberg2016, Gu2017, Roushan2017}. 

\newpage
\section*{\sectionformat V. Conclusions}
We have developed a general theory of time-domain modulation instability of angular momenta in ring array fibers with an arbitrary number of cores. It fully explains the influences of twisting the array on the modulation instability spectra, though we stress that this instability is clearly present in untwisted arrays. In every $N$ core array with fixed material and coupling parameters, there are always $N/2 +1$ distinct MI spectra for $N$ even or $(N+1)/2$ for $N$ odd, due to the relative difference in winding number from the pump. Adjusting the strength of or imparting a Peierls phase to evanescent coupling between fiber cores in the ring allows for a great degree of control over these spectra. This is a promising first step towards wider knowledge of time-domain nonlinear optics in fiber rings, which may enable novel, tunable sources of broadband light carrying angular momentum. 

\bigskip
\textbf{Funding Information:} C.M. acknowledges studentship funding from EPSRC under CM-CDT Grant No. EP/L015110/1. F.B. acknowledges support from the German Max Planck Society for the Advancement of Science (MPG), in particular the IMPP partnership between Scottish Universities and MPG.\\

\bibliographystyle{unsrtnat}
\bibliography{TwistedFiberIdeasBib}

\end{document}